\documentclass[twoside,11pt] {article}

\usepackage{graphics}
\usepackage{amssymb}
\usepackage{amsmath}

\begin{document}

\title{The law of evolution of energy density in the universe
imposed by quantum cosmology and its consequences}
\author{V.E. Kuzmichev and V.V. Kuzmichev\\[0.5cm]
\itshape Bogolyubov Institute for Theoretical Physics,\\
\itshape Nat. Acad. of Sci. of Ukraine, Kiev, 03143 Ukraine}
\date{}

\maketitle

\pagestyle{myheadings} \thispagestyle{plain} \markboth{V.E.
Kuzmichev and V.V. Kuzmichev}{The law of evolution of energy
density} \setcounter{page}{1}

\begin{abstract}
The quantum model of homogeneous and isotropic universe filled
with the uniform scalar field is considered. The time-independent
equation for the wavefunction has the solutions which describe the
universe in quasistationary states. The evolution of the universe
is realized in the form of transitions between such states. In the
first stage in the early universe scalar field slow rolls into a
vacuum-like state with a vanishing energy density. In the second
stage this field begins to oscillate near the minimum of its
potential energy density and becomes a source of creation of
matter/energy in the universe. The quantum model predicts
effective inverse square-law dependence of the mean total energy
density $\overline{\rho}$ on the expectation value of cosmological
scale factor $\langle a \rangle$ where the averaging is performed
over the state with large quantum numbers. Such a law of
decreasing of $\overline{\rho}$ during the expansion of the
universe allows to describe the observed coordinate distances to
type Ia supernovae and radio galaxies in the redshift interval $z
= 0.01 - 1.8$. A comparison with phenomenological models with the
cosmological constant ($\Lambda$CDM) and with zero dark energy
component ($\Omega_{M} = 1$) is made. It is shown that observed
small deviations of the coordinate distances to some sources from
the predictions of above mentioned simple quantum model can be
explained by the fluctuations $\delta a$ of the scale factor about
the average value $\langle a \rangle$. These fluctuations can
arise due to finite widths of quasistationary states in the early
universe. During expansion the fluctuations $\delta a$ grow with
time and manifest themselves in the form of observed relative
increase or decrease of coordinate distances. The amplitudes of
fluctuations $\delta a/\langle a \rangle$ calculated from observed
positions of individual supernovae are in good agreement with
their estimations in quantum theory. Proportionality of the
average value $\langle a \rangle$ to total quantity of
matter/energy in the universe on the one hand and to its age on
the other hand predicted by the quantum model agrees with the
present-day cosmological observables. Possible consequences from
the conclusions of quantum theory are discussed.
\end{abstract}

\section{Introduction}
\label{Intro}

The observed faintness of the type Ia supernovae (SNe) at the high
redshift \cite{Rie,Per} attracts cosmologists' attention in
connection with the hypothesis of an accelerating expansion of the
present-day universe proposed for its explanation
\cite{Rie,Per,Tu1,Pee,Liv}. Such a conclusion assumes that dimming
of the SNe Ia is hardly caused by physical phenomena non-related
to overall expansion of the universe as a whole, such as
unexpected luminosity evolution, effects of contaminant gray
intergalactic dust, gravitational lensing, or selection biases
(see Ref.~\cite{Ton} for review and \cite{Vish} about the
absorption of light by metallic dust). Furthermore it is supposed
that matter component of energy density in the universe
$\rho_{M}$, which includes visible and invisible (dark) baryons
and dark matter, varies with the expansion of the universe as
$a^{-3}$ (i.e. it has practically vanishing pressure, $p_{M}
\approx 0$), where $a$ is the cosmological scale factor, while
mysterious cosmic fluid (so-called dark energy \cite{Tu2,Ost}) is
describes by the equation of state $p_{X} = w_{X} \rho_{X}$, where
$-1 \leq w_{X} \leq - \frac{1}{3}$ \cite{Tu1,Pee}. Parameter
$w_{X}$ can be constant, as e.g. in the models with the
cosmological constant, $w_{X} = - 1$, ($\Lambda$CDM-models)
\cite{Tu1,Pee,Ton}, or may vary with time as in the rolling scalar
field scenario (models with quintessence) \cite{Pee,Liv}. Even if
regarding baryon component one can assume that it decreases as
$a^{-3}$ (pressure of baryons may be neglected due to their
relative small amount in the universe), for dark matter (whose
nature and properties can be extracted only from its gravitational
action on ordinary matter) such a dependence on the scale factor
may not hold in the universe taken as a whole (in contrast to
local manifestations e.g. in large-scale structure formation,
where dependence $a^{-3}$ may survive). Since the contribution
from all baryons into the total energy density does not exceed 4\%
\cite{Hag}, the evolution of the universe as a whole is determined
mainly by the properties of dark matter and dark energy. The
models of dark energy \cite{Tu1,Pee,Tu2} show explicitly unusual
behaviour of this component during the expansion of the universe.

According to modern astrophysical data the mean energy density in
the present-day universe is estimated as $\rho_{0} \sim 10^{-29}\
\mbox{g cm}^{-3}$ \cite{Hag}, the mass of its observed part is
$M_{0} \sim 10^{80}$ GeV, while its radius of curvature is $a_{0}
\sim 10^{28}$ cm \cite{MTW,DZS,Lin}. The age of the universe
equals to $t_{0} \sim 10^{17}$ s \cite{Pee,Hag,Kra}. At the same
time the dimensionless age parameter is $H_{0}\,t_{0} \sim 1$,
where $H_{0}$ is the present-day value of the Hubble constant
\cite{Pee,Ton}. If one expresses these values in modified Planck
units, where length and density are measured in $l_{P} =
\sqrt{2G/(3\pi)}$ and $\rho_{P} = 3/(8 \pi G l_{P}^{2})$
respectively \cite{K,KK}, the simple relation between the observed
parameters of the present-day universe will be revealed
\begin{equation}\label{1}
    M_{0} \sim a_{0} \sim t_{0} \sim 10^{61},
\end{equation}
while its total energy density will be
\begin{equation}\label{2}
    \rho_{0} \sim \frac{1}{a_{0}^{2}} \sim \frac{1}{t_{0}^{2}} \sim
    10^{-122}.
\end{equation}
If these relations will not be considered as an accidental
coincidence realized for unknown reason at the present epoch (as
in the hypothesis of fine tuning or compensation of the
cosmological constant \cite{Pee,Wei}), it is reasonable to assume
the existence of the epoch, when the energy density in the
universe effectively decreases as $a^{-2}$. In the present paper
we pay attention to the fact that exactly such a dependence of the
mean energy density $\overline{\rho}$ on the average $\langle a
\rangle$ in the state with large quantum numbers which describes
only homogenized properties of the universe is predicted by the
quantum model of the homogeneous and isotropic universe proposed
in Refs.~\cite{K,KK}. We show that the quantum model, where the
density $\overline{\rho} \sim \langle a \rangle ^{-2}$, allows to
explain the observed coordinate distances to SNe Ia and radio
galaxies (RGs) in wide redshift range $z = 0.01 - 1.8$. The
observed small deviations of the coordinate distances to some
sources from the predictions of the model with $\overline{\rho}
\sim \langle a \rangle ^{-2}$ can be explained by the local
manifestations of quantum fluctuations $\delta a$ of the scale
factor about its average value $\langle a \rangle$. These
fluctuations produce accelerating or decelerating expansions of
space subdomains containing separate sources with high redshift
whereas the universe as a whole expands at a steady rate. The
amplitudes of fluctuations $\delta a/\langle a \rangle$ calculated
from observed positions of individual supernovae are in good
agreement with their estimations in quantum theory. We make a
comparison with phenomenological $\Lambda$CDM-model and the model
with zero dark energy component ($\Omega_{M} = 1$). Possible
consequences from the conclusions of quantum theory are discussed.

\section{Quantum model}
\label{Quant}

Just as in ordinary quantum nonrelativistic and relativistic
theories one can assume that the problem of evolution and
properties of the universe as a whole in quantum cosmology should
be reduced to the solution of the functional partial differential
equation determining the eigenvalues and the eigenstates (in space
of generalized variables, whose roles are played by the metric
tensor components and matter fields) of some hamiltonian-like
operator. In such an approach the Wheeler-DeWitt equation
\cite{Whe,DeW} will be the equation with zero eigenvalue. As it is
shown in Refs.~\cite{K,KK} the homogeneous, isotropic and
spatially closed universe filled with primordial matter in the
form of the uniform scalar field $\phi$ with some potential energy
density $V(\phi)$ is described by the equation (all in units of
$l_{P} =1$, $\rho_{P} =1$)
\begin{equation}\label{3}
 \left( -\,\partial _{a}^{2} +  a^{2} - a^{4} \hat{\rho}_{\phi}  - E  \right)  \psi _{E} = 0,
\end{equation}
where the operator
\begin{equation}\label{4}
\hat{\rho}_{\phi} = -\, \frac{2}{a^{6}}\,\partial _{\phi }^{2} +
V(\phi),
\end{equation}
corresponds to the energy density of the scalar field in classical
theory (cf. e.g. Ref.~\cite{Pee}), while the wavefunction $\psi
_{E}$ is given in $(a,\phi)$-space of two variables, the scale
factor $a$ and matter field $\phi$. The eigenvalue $E$ determines
the components of the energy-momentum tensor
\begin{equation}\label{5}
  \widetilde T^{0}_{0} = \frac{E}{a^{4}},\quad
 \widetilde T^{1}_{1} = \widetilde T^{2}_{2} = \widetilde T^{3}_{3} =
 -\,\frac{E}{3\, a^{4}},\quad \widetilde T^{\mu }_{\nu } = 0
  \ \ \mbox{for} \ \  \mu \neq \nu.
\end{equation}
For $E > 0$ it coincides with the energy-momentum tensor of
relativistic matter. But for $E < 0$ one cannot associate any
physical source with tensor $\widetilde T^{\mu }_{\nu }$. We shall
consider the case $E > 0$ and call a source determined by the
energy-momentum tensor (\ref{5}) a radiation.

From Eq.~(\ref{3}) it follows the relation for the average values
in the state $\psi _{E}$ normalized in one way or another
\begin{equation}\label{6}
 \left \langle -\,\frac{1}{a^{4}}\,\partial _{a}^{2} \right \rangle =
 \left \langle \hat{\rho}_{\phi}\right \rangle + \left \langle \frac{E}{a^{4}}\right \rangle -
 \left \langle \frac{1}{a^{2}} \right \rangle.
\end{equation}
We shall assume that the average value $\langle a \rangle$
determines the scale factor of the universe in classical
approximation. Then the relation (\ref{6}) takes the form of the
first Einstein-Friedmann equation in terms of average values (see
Appendix \ref{A})
\begin{equation}\label{7}
    \left(\frac{1}{\langle a \rangle}\,\frac{d \langle a
    \rangle}{dt}\right)^{2} =  \overline{\rho}  - \frac{1}{\langle a
    \rangle^{2}},
\end{equation}
where
\begin{equation}\label{8}
    \overline{\rho} = \frac{2}{\langle a \rangle ^{6}} \left \langle -\,\partial_{\phi}^{2} \right
    \rangle + \left \langle V \right \rangle + \frac{E}{\langle a \rangle ^{4}}
\end{equation}
is the mean total energy density, $H = (1/\langle a \rangle)\,d
\langle a \rangle/dt$ is the Hubble constant.

In order to specify the solution of Eq.~(\ref{3}) at given
$V(\phi)$, it has to be supplemented by boundary conditions. In
the asymptotic region $a^{2} \gg 1$ the solution of (\ref{3}) can
be represented in the form \cite{KK}
\begin{equation}\label{9}
\psi_{E} \sim c^{(-)}(E)\, \varphi^{(-)}_{E} + c^{(+)}(E)\,
\varphi^{(+)}_{E},
\end{equation}
where $\varphi^{(-)}_{E}(a, \phi)$ and $\varphi^{(+)}_{E}(a,
\phi)$ are the wave incident upon the barrier $U = a^{2} - a^{4}
V$ and the outgoing wave respectively, which are considered as the
functions of $a$ at given value of the field $\phi$. The
$c^{(\pm)}(E)$ are some coefficients which depend on $E$. The
boundary condition $c^{(-)}(E) = 0$ selects the outgoing wave from
the superposition (\ref{9}). One can introduce an analog of the
S-matrix $S(E) = -\, c^{(+)}(E)/c^{(-)}(E)$, which will have the
poles in the upper half-plane of the complex plane of $E$ at $E =
E_{n} + i \Gamma_{n}$. These values describe the universe in
$n$-th quasistationary state with the parameters $E_{n} > 0$
(position of the level) and $\Gamma_{n} > 0$ (its width), $n = 0,
1, 2 \ldots $ (number of the state) \cite{K,KK}. In a wide variety
of quantum states of the universe, described by Eq.~(\ref{3}),
quasistationary states are the most interesting, since the
universe in such states can be characterized by the set of
standard cosmological parameters \cite{KK}. The wavefunction of
the quasistationary state as a function of $a$ has a sharp peak
and it is concentrated mainly in the region limited by the barrier
$U$. Therefore following Ref.~\cite{Foc} one can introduce some
approximate function which is equal to exact wavefunction inside
the barrier and vanishes outside it. This function can be
normalized and used in calculations of expectation values. Such an
approximation does not take into account exponentially small
probability of tunneling through the barrier $U$. It is valid for
calculation of observed parameters within the lifetime of the
universe, when the quasistationary states can be considered as
stationary ones with $E = E_{n}$ (cf. e.g. Ref.~\cite{Baz}).

The quantum state of the universe depends on form and value of the
potential $V(\phi)$. Just as in classical cosmology which uses a
model of the slow-roll scalar field \cite{Lin,Lyt} in quantum
theory based on Eq.~(\ref{3}) it makes sense to consider a scalar
field $\phi$ which slowly evolves (in comparison with a large
increase of the average $\langle a \rangle$) into a vacuum-like
state with $V(\phi_{vac}) = 0$ from some initial state
$\phi_{start}$, where $V(\phi_{start}) \sim \rho_{P}$. The latter
condition allows us to consider the evolution of the universe in
time in classical sense. Reaching the state $\phi_{vac}$ the field
$\phi$ begins to oscillate about the equilibrium vacuum value due
to the quantum fluctuations. Here the potential $V(\phi)$ can be
well approximated by the expression
\begin{equation}\label{10}
    V(\phi) = \frac{m^{2}}{2}\left(\phi - \phi_{vac}\right )^{2},
\end{equation}
where $m^{2} = \left ( d^{2} V/d \phi^{2} \right )_{\phi_{vac}} >
0$. The oscillations in such a potential well can be quantized.
The spectrum of energy states of the field $\phi$ obtained here
has a form: $M = m \left (s + \frac{1}{2} \right )$, where $m$ is
a mass (energy) of elementary quantum excitation of the vibrations
of the scalar field, while $s$ counts the number of these
excitations. The value $M$ can be treated as a quantity of
matter/energy in the universe.

\section{Mean energy density}
\label{Mean}

The solution of Eq.~(\ref{3}) for the states of the universe with
large quantum numbers, $n \gg 1$ and $s \gg 1$, has a form
\begin{equation}\label{11}
    \psi_{E} = \varphi_{n}(a)\,f_{ns}(\phi),
    \qquad E = 4 \langle a \rangle \left [\langle a \rangle - M
    \right ],
\end{equation}
where
\begin{equation}\label{12}
    \varphi_{n}(a) = \frac{1}{\sqrt{\langle a \rangle}}\,\cos \left
    (\sqrt{2 N + 1}\,a - \frac{N \pi}{2} \right ),
\end{equation}
\begin{equation}\label{13}
    f_{ns}(\phi) = \left [\frac{1}{3\, \langle (\phi - \phi_{vac})^{2} \rangle}\right ]^{1/4}\,
    \cos \left (\sqrt{M (2 N)^{3/2}}\,(\phi - \phi_{vac}) - \frac{s \pi}{2} \right
    ).
\end{equation}
Here $N = 2 n + 1$ defines the number of elementary quantum
excitations related to the vibrations of geometry with Planck
masses in $n$-th state of the universe \cite{KK2,KK3}, while the
functions $\varphi_{n}(a)$ and $f_{ns}(\phi)$ are normalized by
the conditions
\begin{equation}\label{14}
    \int_{0}^{a_{c}}\!\! da\,\varphi_{n}^{2}(a) = 1, \qquad
    \int_{\phi_{-}}^{\phi_{+}}\!\! d\phi\,f_{ns}^{2}(\phi) = 1,
\end{equation}
where $a_{c} = 2 \langle a \rangle$ and $\phi_{\pm} = \phi_{vac}
\pm \sqrt{3\, \langle (\phi - \phi_{vac})^{2} \rangle}$ are the
classical turning points for the potentials $a^{2}$ and (\ref{10})
respectively. The average values $\langle a \rangle$ and $\langle
(\phi - \phi_{vac})^{2} \rangle$ in the state $\psi_{E}$
(\ref{11}) are equal to
\begin{equation}\label{15}
    \langle a \rangle = \frac{\sqrt{2 N + 1}}{2}, \qquad
    \langle (\phi - \phi_{vac})^{2} \rangle = \frac{M}{6\, m^{2}\, \langle a
    \rangle^{3}}.
\end{equation}
One can see that the average value $\langle a \rangle \gg 1$
corresponds to the state with $n \gg 1$. Using the wavefunction
(\ref{11}) for the energy density with the potential (\ref{10}) we
obtain
\begin{equation}\label{16}
    \overline{\rho} = \gamma \, \frac{M}{\langle a \rangle^{3}} +
    \frac{E}{\langle a \rangle^{4}},
\end{equation}
where the coefficient $\gamma = 193/12$ arises in calculation of
expectation value for the operator of energy density of scalar
field and takes into account its kinetic and potential terms. In
matter dominated universe $M \gg E/(4 \langle a \rangle)$ and from
Eqs.~(\ref{11}) and (\ref{16}) it follows that the quantity of
matter/energy $M$ and the mean energy density $\overline{\rho}$ in
the universe taken as a whole (i.e. in quantum states which
describe only homogenized properties of the universe) satisfy the
relations
\begin{equation}\label{17}
    M = \langle a \rangle, \qquad
    \overline{\rho} = \frac{\gamma }{\langle a \rangle^{2}}
\end{equation}
which agree with the relations (\ref{1}) and (\ref{2}).
Substitution of Eq.~(\ref{17}) into (\ref{7}) leads to the density
parameter $\Omega = \overline{\rho}/H^{2}$, where $H^{2}$
coincides with the critical energy density in dimensionless units
being used, equal to $\Omega = 1.066$. It means that the universe
in highly excited states is spatially flat (to within about
$7$\%). This value of $\Omega$ agrees with existing astrophysical
data for the present-day universe: $\Omega = 1 \pm 0.12$
\cite{Be}, $\Omega = 1.02 \pm 0.06$ \cite{Ne}, $\Omega = 1.04 \pm
0.06$ \cite{Pr}, $\Omega = 0.99 \pm 0.12$ \cite{Si}. For other
values of $\Omega$ see e.g. Ref.~\cite{Ol}.

\section{Coordinate distance to source}
\label{Coor}

From Eqs.~(\ref{7}) and (\ref{17}) we find the Hubble constant as
a function of cosmological redshift $z = a_{0}/\langle a \rangle -
1$
\begin{equation}\label{18}
    H(z) = H_{0}\,(1 + z).
\end{equation}
Let us find the dimensionless coordinate distance $H_{0}\,r(z)$ to
source at redshift $z$, where $r(z)$ is given by
\begin{eqnarray}
    \nonumber
  r(z) &=& a_{0} \, \sin \left (\frac{1}{a_{0}} \int_{0}^{z}\!\! \frac{dz}{H(z)}\right )
  \quad \mbox{for} \quad \Omega > 1,\\
    \label{19}
  r(z) &=& \int_{0}^{z}\!\! \frac{dz}{H(z)} \quad \mbox{for} \quad \Omega = 1,\\
    \nonumber
  r(z) &=& a_{0} \, \sinh \left (\frac{1}{a_{0}} \int_{0}^{z}\!\! \frac{dz}{H(z)}\right )
  \quad \mbox{for} \quad \Omega < 1.
\end{eqnarray}
It is connected with a luminosity distance $d_{L}$ by a simple
relation $r(z) = (1 + z)^{-1}\,d_{L}$ (see
Refs.~\cite{Tu1,Vish,DD,We}). For a flat universe with the Hubble
constant (\ref{18}) the dimensionless coordinate distance obeys
the logarithmic law
\begin{equation}\label{20}
     H_{0}\,r(z) = \ln (1 + z).
\end{equation}

\begin{figure}[ht]
\begin{center}
\includegraphics*{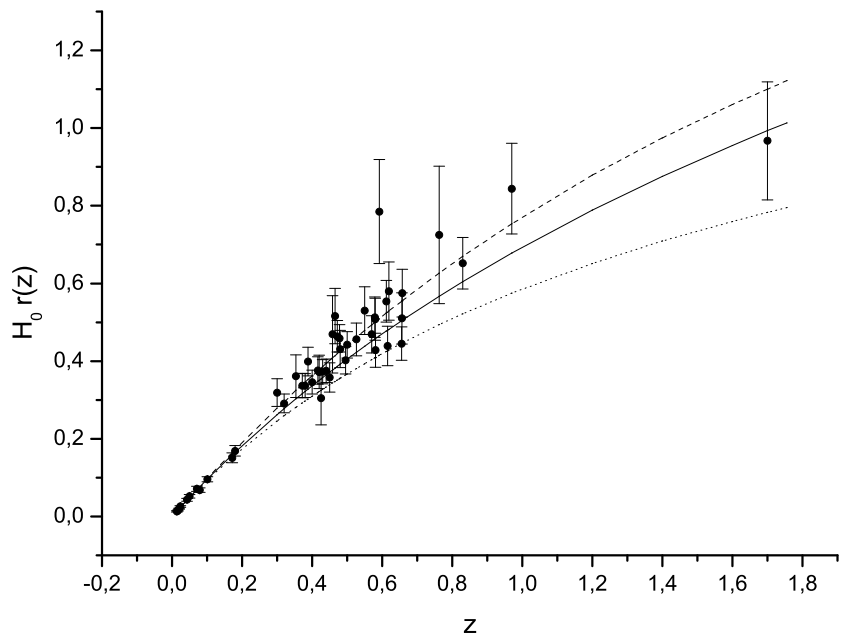}
\end{center}
\caption{Dimensionless coordinate distances $H_{0}\,r(z)$ to
supernovae at redshift $z$. The observed SNe Ia are shown as solid
circles. The model (\ref{20}) is drawn as a solid line. The
$\Lambda$CDM-model with $\Omega_{M} = 0.3$ and $\Omega_{X} = 0.7$
is represented as a dashed line. The model with $\Omega_{M} = 1$
is shown as a dotted line.} \label{fig:1}
\end{figure}

The dimensionless coordinate distances to the SNe Ia and RGs
obtained in Ref.~\cite{DD} from the observational data (solid
circles and boxes) and our result (\ref{20}) (solid line) are
shown in Figs.~1 and 2. The $\Lambda$CDM-model with $\Omega_{M} =
0.3$ (matter component) and $\Omega_{X} = 0.7$ (dark energy in the
form of cosmological constant) and the model without dark energy
($\Omega_{M} = 1$) are drawn for comparison. Among the supernovae
shown in Figs.~1 there are the objects with central values of
coordinate distances which are better described by the
$\Lambda$CDM-model (e.g. 1994am at $z = 0.372$; 1997am at $z =
0.416$; 1995ay at $z = 0.480$; 1997cj at $z = 0.500$; 1997H at $z
= 0.526$; 1997F at $z = 0.580$), the law (\ref{20}) (e.g. 1995aw
at $z = 0.400$; 1997ce at $z = 0.440$; 1995az at $z = 0.450$;
1996ci at $z = 0.495$; 1996cf at $z = 0.570$; 1996ck at $z =
0.656$) and the model with $\Omega_{M} = 1$ (1994G at $z = 0.425$;
1997aj at $z = 0.581$; 1995ax at $z = 0.615$; 1995at at $z =
0.655$). The RG data \cite{DD} demonstrate the efficiency of the
model (\ref{17}) as well (Fig.~2). The quantum model predicts the
coordinate distance to SN 1997ff at $z \sim 1.7$ which is very
close to the observed value (see Fig.~1). In the range $z \leq
0.2$ three above mentioned models give in fact the same result.

The density $\overline{\rho}$ (\ref{16}) contains all possible
matter/energy components in the universe. Let us separate in
(\ref{16}) the baryon matter density equal to $\Omega_{B} \approx
0.04$ \cite{Kra,Ol} which makes a small contribution to the matter
density $\Omega_{M} \approx 0.3$ \cite{Ton}. If we assume that the
baryon density varies as $\rho_{B} \sim a^{-3}$, while the
remaining constituents of density effectively decrease as
$a^{-2}$, then the value $H_{0} r(z)$ calculated in such a model
will practically coincide with the coordinate distance shown in
Fig.~1 as a solid line.

\begin{figure}[ht]
\begin{center}
\includegraphics*{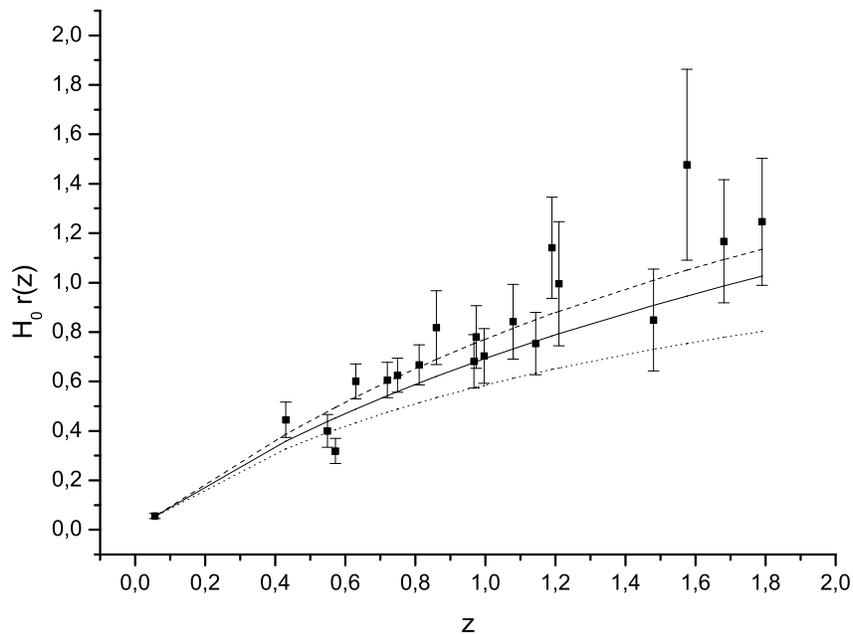}
\end{center}
\caption{Dimensionless coordinate distances $H_{0}\,r(z)$ to radio
galaxies at redshift $z$. Radio galaxies are shown as solid boxes.
The rest as in Fig.~1.} \label{fig:2}
\end{figure}

In Ref.~\cite{Vish} a conclusion is drawn that the model of dark
energy with $w_{X} = - \frac{1}{3}$ implying $\rho_{X} \sim
a^{-2}$ agrees with the recent CMB observations made by WMAP as
well as with the high redshift supernovae Ia data. Such a universe
is decelerating. In our quantum model the total dark matter/energy
in the states which describe only homogenized properties of the
universe varies effectively as $a^{-2}$. In terms of general
relativity it means that its negative pressure compensates for
action of gravitational attraction and the universe as a whole
expands at a steady speed.

\section{Quantum fluctuations of scale factor}
\label{Quan}

Deviations of $H_{0}\,r(z)$ from the law (\ref{20}) towards both
larger and smaller distances for some supernovae can be explained
by the local manifestations of quantum fluctuations of scale
factor about the average value $\langle a \rangle$ which arose in
the Planck epoch ($t \sim 1$) due to finite widths of
quasistationary states. As it is shown in Refs.~\cite{K,KK} such
fluctuations can cause the formation of nonhomogeneities of matter
density which have grown with time into the observed large-scale
structures in the form superclusters and clusters of galaxies,
galaxies themselves etc. Let us consider the influence of
mentioned fluctuations on visible positions of supernovae.

The position of quasistationary state $E_{n}$ can be determined
only approximately, $E_{n} \rightarrow E_{n} + \delta E_{n}$,
where $|\delta E_{n}| \sim \Gamma_{n}$, $\Gamma_{n}$ is the width
of the state. The scale factor of the universe in the $n$-th state
can be found only with uncertainty,
\begin{equation}\label{21}
    \langle a \rangle \rightarrow \langle a \rangle + \delta a,
\end{equation}
where the deviation $\delta a \gtrless 0$ is determined by both
the value $\delta E_{n}$ and the time of its formation
\cite{K,KK}. Since $\Gamma_{n}$ is exponentially small for the
states $n \gg 1$, the fluctuations $\delta E_{n}$ in the early
universe are the main source for $\delta a$. The calculations
demonstrate that the lowest quasistationary state has the
parameters $E_{n=0} = 2.62$ and $\Gamma_{n=0} = 0.31$ (in
dimensionless units). The radius of curvature is $\langle a
\rangle_{n=0} \sim 1$, while the lifetime of such a universe is
$\tau \sim \Gamma_{n=0}^{-1} \sim 3$. Within the time interval
$\Delta t \leq 3$ the nonzero fluctuations of scale factor with
relative deviation equal e.g. to
\begin{eqnarray}
    \nonumber
    \left|\frac{\delta a}{\langle a \rangle}\right| & \lesssim & 0.022
    \quad
    \mbox{at} \ \ \Delta t=1, \\
    \label{22}
    \left|\frac{\delta a}{\langle a \rangle}\right| & \lesssim & 0.040
    \quad
    \mbox{at} \ \ \Delta t=2, \\
    \nonumber
    \left|\frac{\delta a}{\langle a \rangle}\right| & \lesssim & 0.077
    \quad
    \mbox{at} \ \ \Delta t=3
\end{eqnarray}
can be formed in the universe (see Appendix \ref{B}). Such
fluctuations of the scale factor cause in turn the fluctuations of
energy density which can result in formation of structures with
corresponding linear dimensions under the action of gravitational
attraction. For example, for the current value $\langle a \rangle
\sim 10^{28}$ cm the dimensions of large-scale fluctuations
$\delta a \lesssim 70$ Mpc, $\delta a \lesssim 120$ Mpc, and
$\delta a \lesssim 200$ Mpc correspond to relative deviations
(\ref{22}). On the order of magnitude these values agree with the
scale of superclusters of galaxies.

If one assumes that just the fluctuations $\delta a$ cause
deviations of positions of sources at high redshift from the law
(\ref{20}), then it is possible to estimate the values of relative
deviations $\delta a/\langle a \rangle$ from the observed values
$H_{0}\,r(z)$. The fluctuations of scale factor (\ref{21})
generate the changes of coordinate distances,
\begin{equation}\label{23}
    H_{0}\,r(z) = \ln \left[\left(1 +
    \frac{\delta a}{\langle a \rangle} \right)^{-1} (1 + z)\right].
\end{equation}
The possible values of coordinate distances obtained from
Eq.~(\ref{23}) for the relative deviations (\ref{22}) are shown as
a shaded area in Fig.~3. Practically all supernovae in this
redshift interval fall within the limits of (\ref{22}). The only
exception is SN 1997K at $z = 0.592$ which should be characterized
by too sharp negative relative deviation $\delta a / \langle a
\rangle = - 0.274$ (for central value) even in comparison with the
largest possible fluctuations of the scale factor.

\begin{figure}[ht]
\begin{center}
\includegraphics*{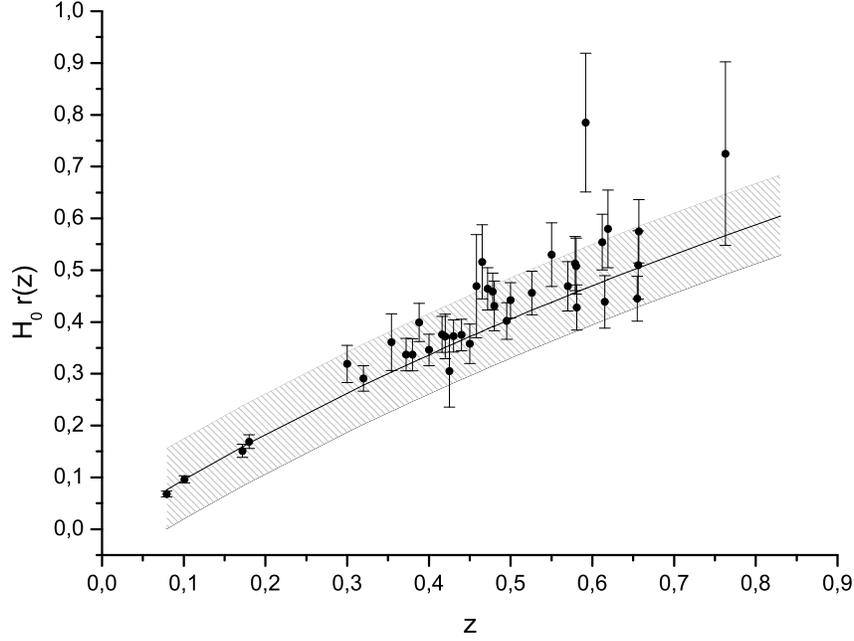}
\end{center}
\caption{Dimensionless coordinate distances $H_{0}\,r(z)$ to
supernovae in the interval $z = 0.1 - 0.8$. A shaded area
corresponds to possible values of coordinate distances in the
model (\ref{23}) for the relative deviations (\ref{22}). The rest
as in Fig.~1.} \label{fig:3}
\end{figure}

The same analysis one can make for RGs as well.

Thus the observed faintness of the SNe Ia can in principle be
explained by the logarithmic-law dependence of coordinate distance
on redshift in generalized form (\ref{23}) which takes into
account the fluctuations of scale factor about its average value.
These fluctuations can arise in the early universe and grow with
time into observed deviations of the coordinate distances of
separate SNe Ia at the high redshift. They produce accelerating or
decelerating expansions of space subdomains containing such
sources whereas the universe as a whole expands at a steady rate.

\section{Some cosmological consequences}
\label{Cosmo}

In matter dominated universe from Eqs.~(\ref{7}) and (\ref{17}) it
follows that
\begin{equation}\label{24}
    \langle a \rangle \sim t,\quad H t = 1
\end{equation}
for any value of $z$. The first relation agrees with the
astrophysical data (\ref{1}) for the present-day universe. The
dimensionless age parameter is known in the range $0.72 \lesssim
H_{0} t_{0} \lesssim 1.17$ with the central value $H_{0} t_{0}
\approx 0.89$ \cite{Pee}. The other values $H_{0} t_{0} = 0.96 \pm
0.04$ \cite{Ton} and $H_{0} t_{0} \approx 0.93$ \cite{Kra} agree
with the theoretical prediction (\ref{24}) as well.

In radiation dominated universe from Eq.~(\ref{7}) at fixed $E$ it
follows the ``standard'' expression for the scale factor (see e.g.
Ref.~\cite{We})
\begin{equation}\label{25}
    \langle a \rangle = \left (2 \sqrt{E} t \right )^{1/2},
\end{equation}
where the time $t$ is counted from the singular state with zero
value of the scale factor. The solution (\ref{25}) is the special
case of the solution given in the Appendix \ref{B}. In general
relativity and in quantum theory the evolution of the universe is
described differently. In general relativity a scale factor is a
function of time, e.g. as in (\ref{25}). The quantum model is
based on the time-independent equation (\ref{3}). And the
evolution is described as successive transitions of the universe
from one state (e.g. with number $n$) to another ($n'$). From the
first equation (\ref{15}) it follows that such transitions will
manifest themselves in the form of expansion ($n' > n$) or
contraction ($n' < n$) of the universe. Direct calculations
\cite{K,KK} and physical arguments \cite{KK2,KK3} show that the $n
\rightarrow n'= n + 1$ transitions are the most probable. As a
result the universe can get into the state with $\langle a \rangle
\sim \sqrt{n} \gg 1$ in a finite time.

In the quantum model every state of the universe is characterized
by its own eigenvalue $E$. In the epoch, when the contribution
from matter/energy in the form of elementary quantum excitations
of the vibrations of the field $\phi$ into the mean total energy
density can be neglected (in comparison with the contribution from
the relativistic matter), $M \ll E/(4 \langle a \rangle)$, the
universe as a whole, according to (\ref{11}), is characterized by
the parameter $E \sim \langle a \rangle ^{2}$ and the energy
density $\overline{\rho} \approx E/\langle a \rangle ^{4}$ will
effectively decrease as $\langle a \rangle ^{-2}$. Then the
solution of Eq.~(\ref{7}) will have the same form (\ref{24}) as in
subsequent matter dominated universe. The same result can be
formally obtained directly from Eq.~(\ref{25}) as well if one
assumes that in accordance with quantum mechanical treatment it
describes the state in which the eigenvalue $E \sim \langle a
\rangle ^{2}$ corresponds to given $\langle a \rangle$. Since in
quantum description after transition from radiation dominated to
matter dominated universe the effective density $\overline{\rho}$
still decreases as $\langle a \rangle ^{-2}$, then the law of
expansion also does not change. The linear dependence of $\langle
a \rangle$ on $t$ as in Eq.~(\ref{24}), if it is realized in the
universe during long enough period of time (possibly with the
exception of short epoch in history of the early universe when
$\langle a \rangle \sim 1$), allows to solve the old problems of
standard cosmology (e.g. the problems of flatness, horizon, and
age \cite{DZS,Lin}) without appeal to the hypothesis about
de-Sitter (exponential) stage of expansion of the early universe
\cite{Lin,Lyt}.

In addition to the prediction about the steady-speed expansion of
the universe as a whole (at the same time the accelerating or
decelerating motions of its subdomains remain possible on a
cosmological scale as it is shown in Sec.~\ref{Quan} of this
paper) the quantum model allows an increase of quantity of
matter/energy in matter dominated universe according to
(\ref{11}). If the mass $m$ of elementary quantum excitations of
the vibrations of the field $\phi$ remains unchanged during the
expansion of the universe, then the increase of $M$ can occur due
to increase in number $s$ of these excitations. But the increase
in $s$ does not mean that a quantity of observed matter in some
chosen volume of the universe increases. According to the model
proposed in Refs.~\cite{KK2,KK3} the observed ``real'' matter
(both luminous and dark) is created as a result of the decay of
elementary quantum excitations of the vibrations of the field
$\phi$ (under the action of gravitational forces) into baryons,
leptons and dark matter. The undecayed part of them forms what can
be called a dark energy. Such a decay scheme leads to realistic
estimates of the percentage of baryons, dark matter and dark
energy in the universe with $\langle a \rangle \gg 1$ and $M \gg
1$. Despite the fact that the quantity of matter/energy can
increase, the mean total energy density decreases and during the
expansion of the universe mainly the number of elementary quantum
excitations of the vibrations of the field $\phi$ increases. Their
decay probability is very small, so that basically only the dark
energy is created. These circumstances can explain the absence of
observed events of creation of a new baryonic matter on a
cosmologically significant scale.

The proposed approach to the explanation of observed dimming of
some SNe Ia may provoke objections in connection with the problem
of large-scale structure formation in the universe, since the
energy density $\overline{\rho}$ in the form (\ref{17}) cannot
ensure an existence of a growing mode of the density contrast
$\delta \overline{\rho}/\overline{\rho}$ (see e.g.
Refs.~\cite{Ol,We,Pee2}). As we have already mentioned above in
Sec.~\ref{Mean} of this paper the density $\overline{\rho}$
(\ref{17}) describes only homogenized properties of the universe
as a whole. It cannot be used in calculations of fluctuations of
energy density about the mean value $\overline{\rho}$. Under the
study of large-scale structure formation one should proceed from
the more general expression for the energy density (\ref{16}).
Defining concretely the contents of matter/energy $M$, as for
instance in the model of creation of matter mentioned above, one
can make calculations of density contrast as a function of
redshift. The problem of large-scale structure formation is one of
the main problems of cosmology (see e.g. Refs.~\cite{Ol,Dol}). It
goes beyond the tasks of this paper and requires a special
investigation. The ways of its solution in the quantum model are
roughly outlined in Ref.~\cite{KK}.

\appendix

\section{Appendix: Derivation of Eq.~(\ref{7})}
\label{A}

\renewcommand{\theequation}{\thesection \arabic{equation}}
\setcounter{equation}{0}

The time-dependent equation which describes the quantum model of
the homogeneous, isotropic and spatially closed universe has a
form \cite{K,KK}
\begin{equation}\label{A1}
    i\,\partial_{T} \Psi = \hat{\mathcal{H}} \Psi,
\end{equation}
where
\begin{equation}\label{A2}
    \hat{\mathcal{H}} = \frac{1}{2} \left(\partial_{a}^{2} -
    \frac{2}{a^{2}}\,\partial_{\phi}^{2} - a^{2} + a^{4} V(\phi)
    \right )
\end{equation}
is a Hamiltonian-like operator. The wavefunction $\Psi$ depends on
a scale factor $a$, scalar field $\phi$, and time coordinate $T$.
In derivation of Eq.~(\ref{A1}) the time $T$ is introduced as an
additional (embedding) variable which describes a motion of a
source in a form of relativistic matter of an arbitrary nature. It
is related to the synchronous proper time $t$ by the differential
equation: $dt = a\, dT$ \cite{K}. Eq.~(\ref{A1}) allows a
particular solution with separable variables
\begin{equation}\label{A3}
    \Psi = \mbox{e}^{\frac{i}{2} E T} \psi_{E},
\end{equation}
where the function $\psi_{E}$ satisfies the time-independent
equation (\ref{3}). The general solution of Eq.~(\ref{A1}) has a
form of the superposition of the states (\ref{A3}) with some
weighting function which characterizes the distribution in $E$ of
the states at the instant $T = 0$ \cite{KK}.

Using Eq.~(\ref{A1}) and taking into account that the Hamiltonian
(\ref{A2}) is Hermitean we obtain the equation which determines a
change in time $T$ of the average value of the physical quantity
$A$
\begin{equation}\label{A4}
    \frac{d}{dT}\langle \hat{A} \rangle =
    \frac{1}{i} \langle [\hat{A},\hat{\mathcal{H}}] \rangle
    + \langle \partial_{T} \hat{A} \rangle,
\end{equation}
where the operator $\hat{A}$ corresponds to $A$,
$[\hat{A},\hat{\mathcal{H}}] = \hat{A} \hat{\mathcal{H}} -
\hat{\mathcal{H}} \hat{A}$, and angle brackets denote the average
value in the state $\Psi$. Introducing the operator $d \hat{A}/dT$
by the relation
\begin{equation}\label{A5}
    \langle \frac{d \hat{A}}{dT} \rangle =
    \frac{d}{dT}\langle \hat{A} \rangle,
\end{equation}
Eq.~(\ref{A4}) can be rewritten in the operator form
\begin{equation}\label{A6}
    \frac{d \hat{A}}{dT} =
    \frac{1}{i} [\hat{A},\hat{\mathcal{H}}]
    + \partial_{T} \hat{A}.
\end{equation}
Setting $\hat{A} = \hat{a}$, and using the explicit form of the
Hamiltonian (\ref{A2}) we find
\begin{equation}\label{A7}
    a\,\frac{d \hat{a}}{d t} = - \hat{\pi}_{a},
\end{equation}
where $\hat{\pi}_{a} = - i\, \partial_{a}$ is the momentum
operator canonically conjugate with $a$, and the operator $\hat{a}
= a$. The operator equation (\ref{A7}) is equivalent to the
definition of the momentum $\pi_{a} = - a\,d a/d t$ canonically
conjugate with the variable $a$ in classical cosmology
\cite{Lin,KK}. Using (\ref{A7}) we obtain
\begin{equation}\label{A8}
    \langle -\, \frac{1}{a^{4}}\, \partial_{a}^{2} \rangle =
    \langle \left (\frac{1}{a}\, \frac{d \hat{a}}{d t} \right )^{2}
    \rangle.
\end{equation}
Let us represent the operators $\hat{a}$ and $d \hat{a}/d t$ as
follows:
\begin{equation}\label{A9}
    \hat{a} = \langle a \rangle + \hat{\xi}, \qquad
    \frac{d \hat{a}}{d t} = \frac{d \langle a \rangle}{d t} +
    \frac{d \hat{\xi'}}{dt},
\end{equation}
where generally speaking the time derivative of the operator
$\hat{\xi}$ is not equal to $d \hat{\xi'}/d t$. Then
\begin{equation}\label{A10}
    \langle \left (\frac{1}{a}\, \frac{d \hat{a}}{d t} \right )^{2}
    \rangle = \langle \left (1 + \frac{\hat{\xi}}{\langle a \rangle}\right )^{-2}
    \left (1 + \frac{d \hat{\xi'}}{d \langle a \rangle}\right )^{2} \rangle
    \left (\frac{1}{\langle a \rangle}\, \frac{d \langle a \rangle}{d t} \right )^{2}.
\end{equation}
In a first approximation one can neglect the deviation of $a$ from
its average value $\langle a \rangle$ and set $a = \langle a
\rangle$. Then using Eqs.~(\ref{A8}) and (\ref{A10}) and taking
into account that $a$ and $\phi$ are independent variables,
Eq.~(\ref{6}) can be reduces to the form (\ref{7}).

Let us note that setting $\hat{A} = \hat{\pi}_{a}$ from
Eq.~(\ref{A6}) we obtain
\begin{equation}\label{A11}
    a\, \frac{d \hat{\pi}_{a}}{dt} = \frac{2}{a^{3}}\,
    \hat{\pi}_{\phi}^{2} + a - 2 a^{3} V(\phi),
\end{equation}
where $\hat{\pi}_{\phi} = - i\, \partial_{\phi}$ is the momentum
operator canonically conjugate with $\phi$. From (\ref{A11}) it
follows the second Einstein-Friedman equation for average values.
Similarly setting $\hat{A} = \phi$ and $\hat{A} =
\hat{\pi}_{\phi}$ one can obtain equations which describe an
evolution in time of the field $\phi$.

\section{Appendix: Fluctuations of scale factor}
\label{B}

\setcounter{equation}{0}

In order to estimate an amplitude of fluctuations of scale factor
(relative deviation $\delta a/\langle a \rangle$) we shall
consider the solution of Eq.~(\ref{7}) for average values in the
epoch when the matter is represented by a slow-roll scalar field
and the kinetic term in density (\ref{8}) can be neglected. In
this case from Eq.~(\ref{7}) we obtain
\begin{eqnarray}
    \nonumber
    \langle a \rangle & = &
    \left \{\frac{1}{2V_{n}} \left [1 + \left
    (2 V_{n} \alpha^{2} - 1 \right ) \cosh \left (2 \sqrt{V_{n}}
    \Delta t \right ) \right ]\right. \\
    \label{B1}
    & + & \left. \sqrt{\frac{E'}{V_{n}}} \sinh
    \left (2 \sqrt{V_{n}}
    \Delta t \right ) \right \}^{1/2},
\end{eqnarray}
where $V_{n} = \langle V \rangle$ depends on a state of field
$\phi$, $E' = E_{n} - \alpha^{2} + \alpha^{4} V_{n}$, and $\Delta
t = t - t_{\mbox{\scriptsize initial}}$. The solution (\ref{B1})
corresponds to the boundary condition $\langle a \rangle = \alpha$
at $t = t_{\mbox{\scriptsize initial}}$. For $2 \sqrt{V_{n}}
\Delta t \ll 1$ we have
\begin{equation}\label{B2}
    \langle a \rangle = \left [\alpha^{2} + 2 \sqrt{E'} \Delta t
    \right ]^{1/2}.
\end{equation}
Keeping main terms only from (\ref{B1}) we find the following
expression for the amplitude of fluctuations
\begin{equation}\label{B3}
    \frac{\delta a}{\langle a \rangle} = \frac{\frac{1}{4} \frac{\delta E'}{E'}}{1 + \frac{1}{2}
    \sqrt{\frac{V_{n}}{E'}} \left [ \left (\alpha^{2} - \frac{1}{V_{n}} \right )
    \tanh \left (\sqrt{V_{n}} \Delta t \right ) + \alpha^{2} \coth \left (\sqrt{V_{n}} \Delta t \right )
    \right ]}.
\end{equation}
For the lowest state with the parameters $E_{n = 0} = 2.62$,
$\delta E' \approx \Gamma_{n = 0} / 2 = 0.16$, $V_{n = 0} = 0.08$,
and $\alpha \approx 1$ \cite{KK} from (\ref{B3}) we find the
values of relative deviations (\ref{22}).

\end{document}